# Quality quantification in Systems Engineering from the Qualimetry Eye

Yann ARGOTTI
LAAS-CNRS, Université de Toulouse
CNRS, INSA
QAPM department
Renault Software Labs
Toulouse, France
yann.argotti@laas.fr

Claude BARON
LAAS-CNRS, Université de Toulouse
CNRS, INSA
Toulouse, France
claude.baron@laas.fr

Phillipe ESTEBAN
LAAS-CNRS, Université de Toulouse
CNRS, UPS
Toulouse, France
philippe.esteban@laas.fr

*Abstract*—Nowadays, quality definition, assessment, control and prediction cannot easily be missed in systems engineering. One common factor among these activities is quality quantification. Therefore, throughout this paper, the authors focus on the problems relating to quality quantification in systems engineering. They first identify the main drawbacks of the current approaches adopted in this domain. They demonstrate how current solutions are not easily repeatable and adaptable across systems and how in most cases, the related standards such as ISO/IEC 25010 or Automotive-SPICE to cite just a few, are not used as they are within companies today. Fortunately, qualimetry, a young science with the purpose of quality quantification, provides the tools to resolve these gaps. To be able to use these tools, the authors propose a synthetic representation of qualimetry and its six pillars, named the "*House of Qualimetry*" and explain the fundamendal aspects of qualimetry. They identify a set of 8 attributes to characterize the design quality model and based on these attributes, propose a new process to design or adapt the quality model. Among these attributes, a new one is introduced to capture and measure the quality model evolution and adaptation aspect: the polymorphism and the polymorphism degree. Finally, the authors consolidate the measurement part thanks to a new measurement process before returning to the benefits of these contributions to systems engineering.

*Keywords—systems engineering, qualimetry, quality model, measure, polymorphism*

## I. Introduction

Quality quantification activity and its usage in decision making is often underestimated and failures on these activities result to non-quality which costs companies 5% total revenue [1]. Moreover, sometimes it happens that the impact related to these failures is even worse with dramatic consequences. We can refer to some well-known examples. On the 15th of April 1912, RMS Titanic sank during its maiden voyage resulting in the loss of 1,523 people [2]. This number could have been greatly reduced if the correct decision was taken during the design phase regarding the waterproof quality of compartments. Over the 1985-1987 period, Therac-25 caused massive radiation overdoses to six patients [3]. The failures were the result of issues in the design and development process. On the 28th of January 1986, the Challenger spacecraft exploded 73 seconds after its ignition, killing all seven crew members [4], with a root cause mainly associated with NASA's company culture and its decision making processes. On the 4th of June 1996, because of an integer overflow linked to the reuse of the same navigation software than Ariane 4, Ariane 5 was self-destructed less than 40 seconds after ignition as a result [5]. Following an unaddressed major defect in airbag, the bankrupt of Takata, an automotive equipment manufacturer, occurred [6] on the 26th of June 2017. In each all of these tragic events, root causes were either uncaught or unaddressed issue(s) in the design, architecture, product, change, decision or development process. A proper quality quantification could have identified these issues leading to corrective action and resolution before it was too late.

Quality quantification, an implicit activity associated with verification and validation processes, is particularly exercised during the quality control part of these two processes [7]. It governs not only the set of relevant quality characteristics, but also how we measure and assess them to ensure that the system that is designed and produced meets its requirements on time. In addition, it gives us the tools to evaluate how well these requirements are met from a quality perspective.

The current techniques used to quantify quality in systems engineering are usually specialized to a specific domain, adapting standards (e.g. CMMI [8] or ISO/IEC 25010 [9]) or latest research achievements. However, that approach is too centric on the object currently under design, development or production and therefore prevents to generalize and benefit immediately from advances on other areas or systems.

In 1968, a new science finally emerged that could generalize the quantification of quality: Qualimetry. This science covers both the theoretical and applied aspects of quality quantification for any domain whether it is technical or non-technical. Unfortunately, this relatively young science, which has a large scope, is not widely used even in systems engineering where we encounter only specific applied qualimetry case studies which are mostly de-correlated from theoretical qualimetry. Thus, we are proposing to bring this science into systems engineering, showing the field of perspective offered by qualimetry.

In the following sections of this paper, we first review the current context and problems linked to quality quantification in systems engineering and see how qualimetry enables their resolutions. We then propose a synthetized view on qualimetry, represented by what we call the "*House of Qualimetry*", that fosters its understanding, depicting quality models and measurement concepts. Next, we consolidate these two concepts of model and measurement by proposing a unified quality model conception and a new measurement process. Finally, we review the interests, with respect to systems engineering, of a qualimetry approach reinforced with our contributions versus the traditional way of quantifying quality.

## II. CONTEXT AND PROBLEMS OF QUALITY QUANTIFICATION IN SYSTEMS ENGINEERING

We have seen from the examples in the introduction how essential it is to properly evaluate and assess the quality of a system. However, that task is more complex than simply expressing the above sentence. Indeed, it requires that we have a clear definition of: what is beneath quality, the system we aim to evaluate, the system dependencies, the way we are characterizing the system quality, how and when we are measuring these characteristics and controlling then the quality level during each steps of the system life cycle, described in ISO/IEC/IEEE 15288:2015[9]. Hopefully, we have many years of work and literature upon which we can rely.

Starting with quality, while its definition evolved from the 5[th] century B.C. Greek philosopher's thoughts in their quest to know *"what is knowledge?"* [11]–[13], quality of something or someone represents the properties or characteristics of that thing or being. Nowadays the definition is more nuanced in that quality is perceived as positive by default. However, to avoid any further debate about the meaning of quality, we are taking the definition from IEEE Standard glossary [14], which is also the one used in the International Software Testing Qualification Board glossary [15]:

*"The degree to which a system, component, or process meets*

1- *Specified requirements,*

2- *Customer or user needs or expectations"*

Regarding system and system life cycle, INCOSE Handbook [16] gives us the right core knowledge and foundation. Nevertheless, one important thing we have to take into consideration is that a system is a combination of three dimensions: physical, computational (or logical) and human [17]. Therefore, quantifying and then controlling the quality of a system consists of being able to characterize, measure and assess each of these dimensions and their respective combinations.

When we speak about system quality characterization, we refer to the general approach which identifies and organizes over a quality model [12], [18] the relevant characteristics of the system that we have to measure and assess in order to be able to draw a conclusion about its quality level. In 2011, ISO/IEC 25010 standard [10], an evolution and extension to systems engineering field of the previous standard ISO/IEC 9126 [19], was published and is the current reference. Moreover, ISO/IEC 25010 is itself part of the ISO/IEC 250xx standard series [9], [20]–[29] called System and Software Quality Requirements and Evaluations *aka* SQuARE. Interestingly, we are noticing that SQuARE scope does not cover System, Software and Hardware (i.e. physical dimension of a system) like ISO 26262 [30], the *"Road vehicles – Functional safety"* standard does.

However, despite the fact that this standard set gives us three quality models[1] with a focus on system and software, it is a weak standard that requires complements and clarifications. As B. Boehm stated in a recent *Systems Engineering Research Center* talk [31], this standard is too generic in that it attempts to fit each and every case into one and does not consider the evolution of stakeholders needs depending on time, environment and the type of stakeholder. This statement is illustrated in the survey carried out by Wagner et al. [32] who performed a survey focused on practitioners and companies located in German-speaking countries. The result was that quality related standards[2] were used in less than 28% of the cases and out of this 28%, almost 79% were more or less deep adaptations of the quality model specified by these standards. So, our quality quantification dilemma becomes a question of whether it is better to create a new quality model or tailor the existing model.

Most common and early methods to design quality model are the Factor / Criteria / Metric (FCM) by McCall *et al.* in 1977 [33] and its generalization by Basili *et al.* in 1994 [34] into the Goal / Question / Metric (GQM) methods. These methods consist of quality model construction, with corresponding metrics, by answering the questions such as *"what are the system quality factors and their respective criteria?"* or *"what are the system quality goals and their corresponding questions?"*. Many of the quality models have been designed with these methods, including ISO/IEC 25010 which is based on GQM. Unfortunately, these methods are missing some important aspects in the design and adaptation methodology. We can cite for instance, the integration of quality model purpose with the Definition - Assessment - Prediction (DAP) classification from Deissenboeck *et al.* [35], see Fig. 1. This classification depicts the incremental relationship between these three purposes, starting from definition models. Other examples can be that factors or quality characteristics can have different impact, or weight, in the overall system quality, or also, the question of *"among the large number of existing quality models and factors, or quality characteristics, how to select and adapt to them to our system?"*.

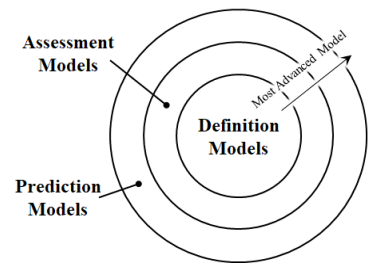

Fig. 1 - The DAP classification introduced by Deissenboeck *et al.* **[35]**

Iqbal and Babar [36] were giving an approach using fuzzy logic to identify which of the ISO/IEC 25010 product quality model characteristics has to be used in their decision support system applied to "*Internet Banking*" case study, and then relied on Likert scale to help on the quantification aspect. On their side, Gitto *et al.* [37] proposed a methodology based on FCM to design complex system quality model. Unfortunately, in both cases, the focus is restricted to some specific subset of system and these authors missed qualimetry, the science of quantification.

Qualimetry, from the Latin *qualis* "of what kind" and the Greek *μετρεω* "to measure", is the science of quality quantification. It is relatively young: its birth as a new scientific discipline occurred in 1968 [12], [38]. Its origin came from the need to have a generalization of quality quantification over any domains and type of object or being.

---

[1] ISO/IEC 25010 describes product quality model, quality in use model and data quality model.

[2] Standards such as ISO/IEC 9126, ISO/IEC 25000, ISO 9001, CMMI

Naturally, as a science, it is composed of both theoretical and applied disciplines, but due to its youth, qualimetry requires some additions to proceed on systems engineering quality quantification. Firstly, a synthetized view is necessary to foster an understanding and capture its "pillars". We also remarked that the current general methodology approach and algorithm [12] to design quality model can be completed and unified based on the work of Wagner *et al.* [18] and achievement done on current quality model likes ISO/IEC 25010, for example.

A proper quality model is one side of the quality quantification problem. Certainly, the other side concerns the measurement of the quality characteristics and especially all measurement process activities. Additionally, and to be complete here, a qualimetry approach must integrate some missing aspects such as an evaluation plan, measurement record, analyze and reports. ISO/IEC 25040 standard [29] defines a linear evaluation process (see Fig. 2) with the same issue that we have with the rest of ISO/IEC 250*nn* standards: it is not precise enough and therefore requires interpretation and strong complement.

Hopefully we may rely on practical work carried out for software related decision makers by McGarry *et al.* [39] which introduces a process that includes evaluation planning, analysis techniques and measurement information models. We can also find the measurement process introduction carried out by Miller *et al.* [40], which has a scope of systems engineering and the process published by Dekkers *et al.* [41], a US-CERT team on secure software development. All these works must be merged together in order to have a consolidated measurement process. That consolidation also depends on the unified quality model conception highlighted in above paragraph and covered in a later section of this paper.

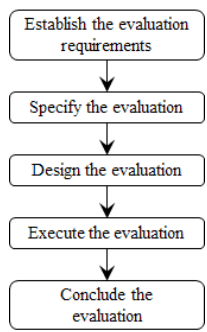

Fig. 2 - Software product quality evaluation process defined by ISO/IEC 25040 **[29]**

Consequently, in the following sections of this paper, we synthetize clearly the different concepts beneath qualimetry to make it more practicable. We then return to the foundation of the quality model design with a focus on qualimetry and propose the conception of a unified quality model which can be applied to any system, even if we integrated some work done within a narrow scope such as software product. We then propose an upgraded measurement process, considering unified conception dependencies and missing parts. Finally, we make a final review and draw a conclusion on the interest of our approach and next steps.

### III. HOUSE OF QUALIMETRY

In order to leverage this science to a large range of audience, foster its accurate understanding and ensure that no major concepts beneath it are eluded or forgotten, we are proposing a synthetized view of the "*House of Qualimetry*" and its 6 pillars, depicted by Fig. 3.

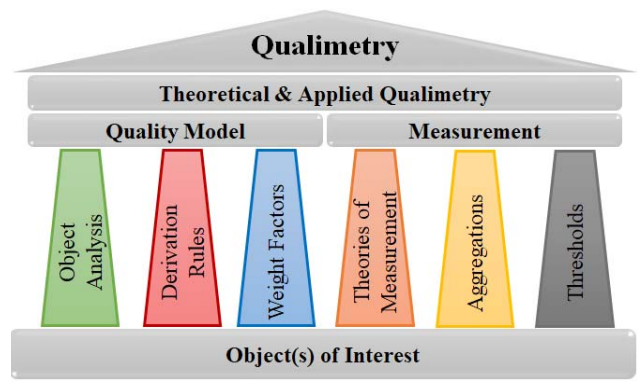

Fig. 3 - The "*House of Qualimetry*" and its 6 pillars.

As a science [3], qualimetry relies naturally on two interlaced and complementary disciplines: theoretical [42] and applied qualimetry [43]. These two disciplines are combined into an entablature which relies on two architraves: "*quality model*" and "*measurement*". "*Quality model*" covers the identification, organization and representation of the relevant quality characteristics while "*measure*" covers the evaluation, manipulation and control of them.

Furthermore, each of these two architraves is relying on a set of three pillars, described in sub-section 1) and 2), settled on a basement reflecting the object(s) of interest (i.e. the one(s) that is (are) aimed to be quality quantified).

*1) "Quality Model" pillars*

While the first pillar (i.e. object analysis) is the major one, the other two are also mandatory in order to achieve the right quality model.

*a) "Object analysis" pillar:* This pillar gathers the necessary knowledge and activities to understand, identify and organize the relevant quality characteristic linked to the analysis of our object of interest (ie the one that its aims to have its quality quantified). Thus, we first define the purpose of our analysis, aligned with the DAP classification (see Fig. 1); we then analyze our object of interest in order to identify the quality characteristics, sub-characteristics and sub-sub-characteristics… that are relevant to us; finally we decide how we are going to organize all this data. We can note that quite often the data organization is achieved via a hierarchical structure (ie tree structure).

*b) "Derivation rules" pillar:* Here, the focus is with regards to global and specific qualimetry rules *[12]* to help optimize the design of the organizational data structure. For example, maximum tree height, division by equal characteristic, branch a tree until only simple or quasi-simple characteristics remain at its top tier.

---
[3] We invite the reader to refer to Azgaldov *et al* [12]. for a demonstration about qualimetry as a science.

*c) "Weight factors" pillar:* Often forgotten, even in standards such as ISO/IEC 25010, the weighting factors are critical because they reflect the importance of quality characteristics among the same level of quality characteristics.

*2) "Measurement" pillars*

As was the case for the previous set of pillars, these thee pillars are all mandatory in order to proceed accurately on measurement taking, even if the "*theories of measurement*" pillar represents the main one.

*a) "Theories of measurement" pillar:* This pillar is composed of three main streams of measurement theories. *[44], [45]*: operational measurement (i.e. how to operate / use the measure), representational measurement (i.e. how to represent the measure) and "various minor" theories. In a sense this is a fundamental pillar as it is bringing together all mathematical and statistical tools for our measurements.

*b) "Aggregations" pillar:* The aim is to deal with the way of combining (i.e. mean, median, variance and more *[46]*) together either all or a subset of the measurements depending on their purpose *[18]*. The aggregated measurements can either be weighted or un-weighted.

*c) "Thresholds" pillar:* This pillar is associated with the measure of the ability to assess, control[4] and therefore make the correct decision. In general, man is using two types of thresholds: acceptance and target. Acceptance is often confused with the reject threshold even though they are not the same: the acceptance threshold is the worst case threshold level that may be accepted, it lies just above the best case reject level. In fact, four types of threshold exist as follows *[12]*: reject, accept, target and reference. Target corresponds to the threshold we are actually aiming for whereas reference corresponds to the reference value used in the industry or in the community at the time when the measurement is taken.

## IV. UNIFIED QUALITY MODEL CONCEPTION

As we have seen in section III, one architrave in the "*House of Quality*" is the quality model. A quality model is an organized and multi-level representation of relevant quality characteristics for an object of interest. The multi-level aspect can be defined as the sub-sequent refinement of characteristics. For example, in ISO/IEC 25010, we have the functional suitability characteristic which is composed of three sub-characteristics: functional completeness, functional correctness and functional appropriateness.

To create such a quality model, we have identified three main streams of approaches, Azgaldov, Wagner and ISO/IEC 250nnn. Azgaldov et al. *[12]* is representing the general qualimetry approach while Wagner *[18]* is describing modeling as an iterative approach, within the software product scope, developed in the Quamoco research project *[47]*. Finally, ISO/IEC 250nnn provides a good illustration of the work that has been done on creating other existing quality models that can be found in literature in general. We may note that there are other works that are very similar to quality model such as McGarry et al. *[39]* but the three above streams are a good synthesis of current distinct approaches. TABLE I provides a summarized comparison between each approach based on their scope and a list of quality model attributes. This attribute list has been elaborated by collecting for each of the three distinct approaches, all the attributes considered by their authors when designing or characterizing the quality model. Indeed, we have noted that even if most of these attributes are in common, they are not equally detailed and used. In addition, and to be complete, we propose an important new attribute, polymorphism. This attribute will be described further in this section.

TABLE I - COMPARISON OF THE MAIN THREE DISTINCT STREAMS OF WORK SUPPORTING QUALITY MODEL SPECIFICATION

| | Stream of approach | Wagner *et al.* [18], [47] | Quality models such as ISO/IEC 250nn [9], [23]–[27], [29] | Azgaldov *et al.* [12], [48] |
|---|---|---|---|---|
| | Quality model scope | Project and Software product | System and Software product and in use | Any area |
| Attributes | 1. *Evaluation context &plan* | none | Evaluation plan | Evaluation context |
| | 2. *Purposes* | • Definition<br>• Assessment<br>• Prediction<br>• Multi-purpose | • Definition<br>• Assessment (evaluation part) | • Definition<br>• Assessment |
| | 3. *QEM: method to assess quality* | Not specified but assumes approximate method | Not specified but assumes approximate method | • Rigorous method<br>• Short-cut method<br>• Approximate method |
| | 4. *QEM: source of information about values in QEM* | Not specified but assumes expert method | Not specified but assumes expert method | • Expert method<br>• Non-expert method (i.e. analytical method)<br>• Hybrid method |
| | 5. *Data organizational types* | • Hierarchical<br>• Meta-model<br>• Statistical and Implicit | • Hierarchical<br>(• Meta-model) | • Hierarchical |
| | 6. *Rules to derives trees* | none | none | ~30 rules |
| | 7. *Weight factors* | Per property / characteristic | Per property/ characteristic | Per property / characteristic |
| | 8. *Polymorphism* | none | none | none |

So, based on these raw comparison results, we first merged them into a consolidated list of attributes to consider when designing or characterizing the quality model. This merge includes the most complete occurrence of each of these attributes. Then, in a second time, we ordered them to generate a unified quality model conception as depicted in Fig. 4: thus, to design or describe a quality model, the user must consider and use sequentially each of these attributes one by one.

*a) Evaluation Context & Plan:* Before performing any further analysis or design of a quality model, the first step is to understand what we want to achieve. For example what is the scope and what are the boundaries of this quality quantification? What are the intentions, limitations, dependencies? What audience are we targeting? What are the responsibilities, timeframe etc…? The answers to these questions provide us with our evaluation context and plan.

---
[4] We can use "criteria" instead of "threshold" particularly for assessment and control, but the concept is identical and "*threshold*" terminology is linked to measurement.

Without knowing them, we won't be able to design the right quality model.

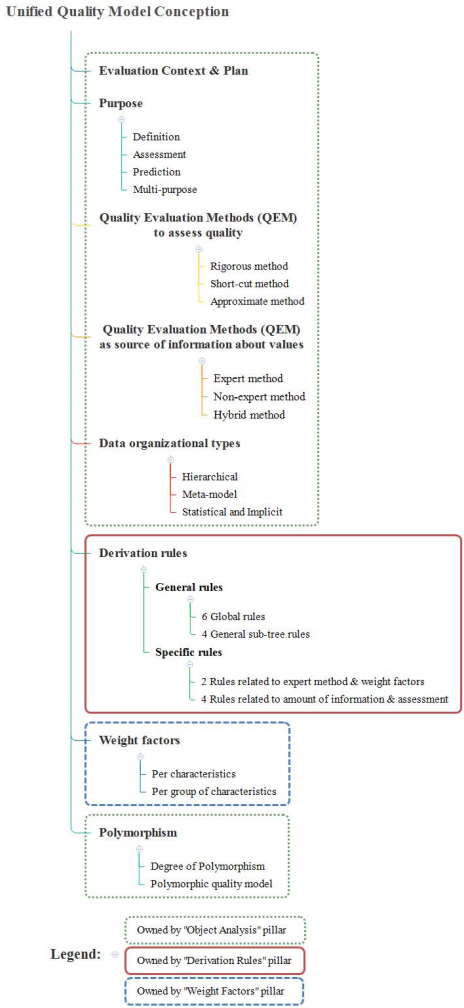

Fig. 4 - The unified quality model conception, aligned with the three "*Quality Model*" pillars

*b) Purpose:* Once we have defined the context, we are in a position to determine the intended usage we are targeting with the quality model being designed. There are three main purposes, the ones described in the DAP classification (see Fig. 1). The Definition purpose corresponds to the description using all the quality characteristics that are relevant and meaningful during quality quantification. The Assessment purpose extends the Definition purpose with corresponding metrics. The Prediction purpose is dedicated to predicting quality. In addition to these 3 categories, we can add a fourth multi-purpose category.. Here the quality model is used, not only for definition but also for assessment and prediction.

*c) Quality Evaluation Methods (QEM):* QEM are not antinomic to FCM and GQM methods. Indeed, while these two methods do help to provide hints on how to find certain quality characteristics[5], QEM are describing two methods linked to how the quality model is going to be designed and subsequently evaluated. The first method characterizes how exhaustive the analysis and quality characterization of our object of interest will be: rigorous method conducts to very detailed quality model while short-cut considers the most essential quality characteristics and therefore leads to a lighter quality model. The second method qualifies the source of information that is being used. It can be based on the findings of domain experts, non-experts or a combination of both.

*d) Data organizational types:* Now that we know context and plan, the purpose of quality model and the QEM to identify our quality characteristics, we have to decide how we are going to organize these data. There are three main types: hierarchical (e.g. tree), meta-model, statistical and implicit. Most of the quality models are taken from the hierarchical type.

*e) Derivation rules:* As we have seen in section III, these rules are guidelines that must be respected during the organization of data. They are composed of global and specific rules mainly dedicated to hierarchical type.

*f) Weight Factors:* This is the same concept as the one described in section III. It is fundamental and must be handled once quality characteristics are identified and organized.

*g) Polymorphism:* We are introducing a new and final attribute to our unified quality model conception: polymorphism. This is the same concept than we have in object-oriented programming. It reflects the capacity of a quality model to describe different types of objects as well as to link with other quality models. To complete this concept we use the nucleotide diversity formula **(1)** introduced by Nei and Li in 1979 *[49]* to measure the degree of polymorphism, or diversity, against other quality models and objects of interest.

$$\pi = \sum_{ij} x_i x_j \pi_{ij} \quad \textbf{(1)}$$

V. MEASUREMENT PROCESS USING QUALITY MODEL

Now that we have set a unified quality model conception to join and extend current quality model design and characterization, we can consider the qualimetry measurement aspect and more particularly the measurement process. Indeed, the aim of a measurement process is not only to proceed on, or collect, measure but also to record and analyze the results, control quality, help on decision making, including doing some predictions and communicating the results to the right stakeholders. If we refer to the current process from ISO/IEC 25040 [29] shown in Fig. 2, we have a coarse and linear definition of the tasks that must be achieved for measurement.

So, as we indicated into section II, we are detailing and completing this process including some practical and complementary work in this field mainly carried out by McGarry *et al.* [39], Miller *et al.* [40], Dekkers *et al.* [41] and Automotive-SPICE[6] [50]. We articulate our proposal of measurement process (cf. Fig. 5) into three sequential phases: Initial, Planning and Execution.

---

[5] We would to raise that both FCM and QCM are not method for quality model design but rather support on how to analyze our object(s) of interest.

[6] For this standard, we are considering MAN.6 management process linked to measurement which gives a set of guidelines that allows us to assess and exercise our proposed measurement process.

*1) Initial phase*

The purpose of this phase is to understand, identify and collect both requirements and context linked to measurement goals and activities. That phase is performed over three tasks which can be realized in parallel. The first task focuses on the identification and enumeration of all measurement objectives, taking measurement requirements as inputs. The second task is dedicated to the measurement context definition which can be understood as defining the scope, the boundaries, the dependencies and the environment linked to measurement activities. The last task of the initial phase relates to process improvement. In the first iteration of these three phases we may not yet have any lessons learned or post-mortem data from previous measurement activities to take into account, however, with time, we will be able to integrate this data in order to improve our current process. The different outputs of these three tasks will be merged and used as inputs to the second phase which is planning.

*2) Planning phase*

During this phase, we transform the requirements, context and process improvement into an evaluation plan, criteria and statistical and/or qualitative techniques to be ready for the execution of that plan. Since that plan must be aligned with the system development life cycle [10], [16], this one is also one input of the planning phase tasks. So, we start to transform measurement requirements and context into the quality model and measurement specifications. Once this has been done, we must plan for their treatment. First by planning for their collection and storage, where processes change, tools and training may be required, then for their analysis procedure and criteria, or thresholds, to apply assessment, control and prediction. The final task of this phase is the synthesis and organization of all outputs from these three previous tasks into one critical document: the evaluation plan.

*3) Execution phase*

The last phase of our process corresponds to the execution of our evaluation plan which is aligned with the system development life cycle phase. The main task here is a loop to collect measurement data at the frequencies defined in the evaluation plan. Each time data is collected, it needs to be stored as well as analyzed and assessed. The results, containing analysis synthesis, predictions, recommendations and conclusions, are generated under various forms - graphical dashboards, analyst summary, detailed results and reports- which are then communicated to the stakeholders, for example, development teams, program managers and any key decision makers.

To conclude on this new process definition achievement, if we do an analogy, for instance, between ISO/IEC 25040 process (cf. Fig. 2) and our proposal (cf. Fig. 5) we can clearly conclude that the standard process is a sub-part of our proposed process. Indeed, "*establish the evaluation requirements*" is a subset of our "*initial phase*", both "*specify the evaluation*" and "*design the evaluation*" are included into our "*planning phase*", and "*execute the evaluation*" with "*conclude the evaluation*" are also a subset of the tasks of our "*execution phase*".

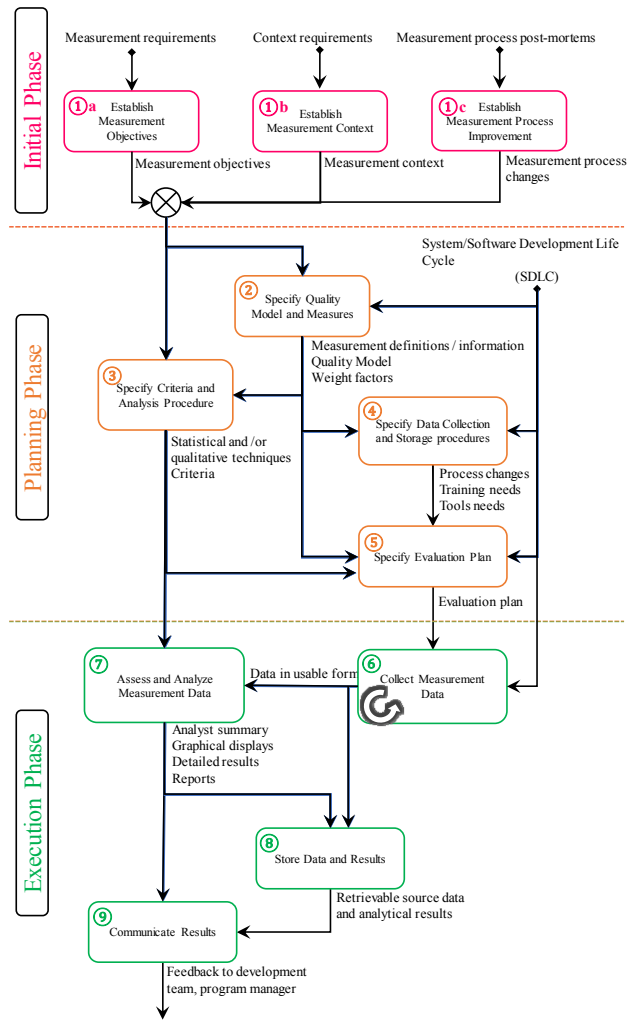

Fig. 5 - Our measurement process proposal articulated over three phases

## VI. INTEREST OF THIS APPROACH

As we have seen, the current solutions are internal and adapted solution with most often some adaptations of standards such as ISO/IEC 250nn which are weak: they are trying to cover everything with only a few quality models, for example. These approaches are exactly the ones that motivated the setup of qualimetry, that it to say, to be able to generalize, adapt and repeat over multiple kinds of system quantification of quality. In other words, we create an approach to quantify quality for a specific object such as a boat or a chair for example, but once we are willing to build another slightly different type of system or object, our quality quantification method often does not fit. So qualimetry is the science that can help us here because its scope is general, repeatable and it gives us theoretical tools to address our needs.

However, this science is still quite young and not well understood. Indeed, we note that in general, like in systems engineering, there is some confusion about the terminology because the term qualimetry is quite often used to describe a specific application on quality measurement and assessment/control quality and therefore belongs to applied qualimetry. So, with our "*House of Qualimetry*", we give a synthetic view of what is behind that science, ensuring that we do not forget to consider the most important aspects during quality quantification. This is what we call the pillars of quality model and measurement.

Moreover, we have proposed a unified design conception for quality model, applicable to any field. It shows that if we use one of the existing quality models, there are quality model attributes that are often forgotten: weight factor aspect is one such attribute, respect of derivation rules is another. Additionally, we have extended the current definition of quality model by introducing the polymorphism concept which captures the fact that quality model can cover multiple types of object and a quality model can be an instantiation of another quality model. This is important because it provides consistency over quality models.

Regarding the measure aspect of qualimetry, we addressed the gaps in current existing solutions and standards when we attempted to apply them to an internal automotive project relying on Automotive-SPICE [50] and more particularly with the MAN.6 measurement management process. A measurement process is obviously not a linear process composed of a few tasks because of its dependencies and the variety of tasks that must be performed. Our proposed process ensures not only that we are identifying each of the measurement requirements and context but also that we are integrating the use of the quality model, record and consumption (i.e. assessment, analysis, prediction, production of reports and dashboards) of measurement data as well as communication to help decision makers.

In a more concrete way and to see how applicable our approach is to real systems, we may consider the automotive field[7]. Indeed, there exists a wide variety of car platforms (e.g. mini-compact, crossover, supercar, convertible, commercial, sport, van ….) that can be considered as variants of a vehicle. Moreover, each type of car platform is a complex system, itself composed of more than 40 systems that are distributed over more than 60 Electronic Control Units (ECU). An ECU is a compound of hardware and software; it is characterized by a set of common characteristics shared with other ECUs (e.g. diagnostic, connection interface, power), a set of specific characteristics (e.g. HMI, communication, safety) and a context (e.g. door control, engine control, telematic control, seat control). Thus, for such complex systems, including all sub-systems, the interest of our approach is that it brings homogeneity, consistency and compatibility to quality quantification. In addition, for the entire complex system - including its different systems-, our approach helps specify a joint "vocabulary", defining a derivable quality model (e.g. ECU or car platform one) and likewise allows smooth incremental change management which is key in agile development methodology

## VII. CONLCUSION

As we observed in the introduction, quantifying quality is key in order to properly assess and control system quality, as well as to provide useful support and data to decision makers. Consequently, this paper has focused on strengthening quality quantification for systems engineering, starting with the main gaps identified.

We have seen that current quality quantification in systems engineering can be consolidated into a specific applied qualimetry case study, which is limited and often prevents replication or generalization in other systems. Moreover, and as Wagner *et al.* survey [32] highlighted, approximately 94% of companies are designing their own quality models diverging more or less from existing standards, such as ISO/IEC 25010 or A-SPICE [50] for instance. The main reason for that divergence is that those models are not sufficiently precise to fit company's needs.

Therefore, we proposed to step back to the foundation and use qualimetry, which is the science of quality quantification, to support us in filling these gaps. Thereby, our first contribution aims to clarify, leverage and foster knowledge related to qualimetry by proposing the synthetic view of the "*House of Qualimetry*" and its six pillars. Then, to support this synthetic view, we have elaborated its two architraves: quality model and measure.

Our second and third contributions were the identification of the height required attributes to characterized and design quality model, and the unified quality model conception (cf. Fig. 4) respectively. This unified conception is a sequential process to design, adapt or replicate quality model.

Moreover, one of these height attributes constitutes our fourth contribution. This is the polymorphism concept applied to our quality model. It captures quality model evolution, adaptation and replication aspects. We completed it with the polymorphism degree which gives us a formula to evaluate intrinsic distance between quality models.

Finally, our last contribution, our measurement process proposal, consolidated the "*House of Qualimetry*", not only by exploiting the "*measure*" architrave, but also establishing a clear link with the "*quality model*" architrave.

In conclusion, this paper opens a new perspective with regards to quality quantification in systems engineering thanks to qualimetry science which gives us the hindsight to fill in the identified gaps using practical solutions.

ACKNOWLEDGMENT

We would like to thank our colleagues at Renault and Renault Software Labs with regards to some fruitful discussions related to quality models and processes linked to measurement. These exchanges were carried out under the scope of an internal Renault-Nissan-Mitsubishi alliance project related to Automotive-SPICE [50].

---

[7] In the same way, we could also take concrete examples from the aeronautical field.